\documentclass[journal]{IEEEtran}
\usepackage{graphicx}

\usepackage{bm}
\usepackage{amsfonts}

\usepackage{amssymb}

\usepackage{color}

\usepackage{cite}

\ifCLASSINFOpdf
\else
\fi
\usepackage{amsmath}
\usepackage{algorithm}
\usepackage{algorithmic}
\ifCLASSOPTIONcompsoc
 \usepackage[caption=false,font=normalsize,labelfont=sf,textfont=sf]{subfig}
\else
 \usepackage[caption=false,font=footnotesize]{subfig}
\fi
\begin{document}
%
\title{On Range Sidelobe Reduction for Dual-functional Radar-Communication Waveforms}
%
%
%

\author{Fan Liu,~\IEEEmembership{Member,~IEEE,}
        Christos Masouros,~\IEEEmembership{Senior~Member,~IEEE,} \\
        Tharmalingam Ratnarajah,~\IEEEmembership{Senior~Member,~IEEE,}
        and~Athina Petropulu,~\IEEEmembership{Fellow,~IEEE,}
\thanks{This work was supported in part by the Marie Sk{\l}odowska-Curie Individual Fellowship under Grant No. 793345, in part by the Engineering and Physical Sciences Research Council (EPSRC) of the UK Grant number EP/S026622/1, and in part by the UK MOD University Defence Research Collaboration (UDRC) in Signal Processing.}
\thanks{F. Liu and C. Masouros are with the Department of Electronic and Electrical Engineering, University College London, London, WC1E 7JE, UK (e-mail: fan.liu@ucl.ac.uk, chris.masouros@ieee.org).}
\thanks{T. Ratnarajah is with the Institute for Digital Communications, School of Engineering, The University of Edinburgh, Edinburgh, EH9 3JL, UK (e-mail: t.ratnarajah@ed.ac.uk).}
\thanks{A. Petropulu is with the Department of Electrical and Computer Engineering, Rutgers, the State University of New Jersey, 94 Brett Road, Piscataway, NJ 08854, United States (e-mail: athinap@rutgers.edu).}
}

\maketitle

\begin{abstract}
In this paper, we propose a novel waveform design for multi-input multi-output (MIMO) dual-functional radar-communication systems by taking the range sidelobe control into consideration. In particular, we focus on optimizing the weighted summation of communication and radar metrics under per-antenna power budget. While the formulated optimization problem is non-convex, we develop a first-order descent algorithm by exploiting the manifold structure of its feasible region, which finds a near-optimal solution within a low computational overhead. Numerical results show that the proposed waveform design outperforms the conventional techniques by improving the communication rate while reducing the range sidelobe level.
\end{abstract}


%
\IEEEpeerreviewmaketitle

\section{Introduction}
%
%
%
%
\IEEEPARstart{T}{O} ease the ever-increasing competition over the scarce spectrum resources, frequency bands currently assigned exclusively to radar systems are expected to be opened up for use by future wireless communication systems. In many emerging applications such as vehicle-to-everything (V2X) networks, it is desirable to have both sensing and communication functionalities over not only the same frequency band, but also on the same hardware platform. Such dual-functional radar-communication (DFRC) systems have attracted a lot of recent research interets \cite{7782415,7347464,8288677,8386661}.
\\\indent One of the most critical challenges in DFRC systems is the design of a joint waveform for simultaneous target detection and communication. Existing contributions aim at designing MIMO DFRC waveforms by using the spatial sidelobes of the transmit beampattern for communication, and the mainlobe for target detection \cite{7347464}. However, such designs are not well-suited in multi-path environments, where the communication symbol received will be masked by the dispersed signals arriving from non-line-of-sight (NLoS) paths. To overcome this drawback, the authors of \cite{8288677} proposed a beamforming design for jointly generating the probing beampattern while guaranteeing the downlink quality-of-service (QoS) of NLoS communication. More relevant to this work, several DFRC waveform designs have been proposed in \cite{8386661} for minimizing the multi-user interference (MUI) in the NLoS channel under given radar-specific constraints, which have achieved favorable performance trade-off between radar and communication. While the above works have realized the basic dual functionality, none of them has considered the range sidelobes in the waveform design, which is a very crucial performance metric for the radar \cite{li2008mimo}. In fact, it is rather difficult to control the range sidelobe level of the DFRC waveform due to the randomness in the communication data embedded. To the best of our knowledge, the above topic remains widely unexplored in the existing DFRC works.
\\\indent In this paper, we extend the work of \cite{8386661} by considering the minimization of range sidelobes, i.e., the time-domain cross-correlation, for MIMO DFRC systems. First of all, we review the closed-form DFRC waveform design proposed by \cite{8386661} without constraining the range sidelobes. As a step further, we incorporate the integrated sidelobe level (ISL) in the objective function, and minimize the weighted summation of the MUI, the Euclidean distance between the designed and the reference waveforms, as well as the ISL under per-antenna power constraint. As the formulated optimization problem is non-convex, we develop a first-order descent algorithm based on the manifold optimization framework, which is able to obtain a near-optimal solution for the problem. Finally, we validate the performance of the proposed waveform design via numerical simulations, showing that the proposed method outperforms the closed-form designs in \cite{8386661} by reducing the range sidelobe level while significantly improving the communication rate.

\section{System Model}
As shown in Fig. 1, we consider a DFRC base station (BS) equipped with an \emph{N}-antenna uniform linear array (ULA), which serves \emph{K} single-antenna users in the downlink while sensing the targets. Below we briefly introduce the mathematical models for both communication and radar operations.
\begin{figure}[!t]
    \centering
    \includegraphics[width=0.6\columnwidth]{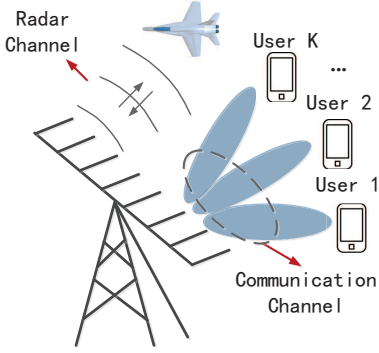}
    \caption{MIMO dual-functional radar-communication system.}
    \label{fig:1}
\end{figure}
\subsection{Communication Model}
Let us consider the transmission of a single DFRC signal block. The received signal matrix at the users is obtained in the form
\begin{equation}\label{eq1}
  {{\mathbf{Y}}}_C = {\mathbf{H}}{\mathbf{X}} + {\mathbf{W}}_C,
\end{equation}
where $\mathbf{H} \in {\mathbb{C}^{K \times N}}$ denotes the communication channel matrix, which is assumed to be perfectly known at the BS, $\mathbf{X} \in {\mathbb{C}^{N \times L}}$ is the DFRC waveform matrix to be designed, with a block length of \emph{L}, and ${\mathbf{W}}_C \in {\mathbb{C}^{K \times L}}$ denotes additive white Gaussian noise (AWGN) with a variance $N_0$.
\\\indent By denoting the symbol matrix sent to \emph{K} users as $\mathbf{S} \in {\mathbb{C}^{K \times L}}$, eq. (\ref{eq1}) can be equivalently expressed as
\begin{equation}\label{eq2}
    {{\mathbf{Y}}}_C = {\mathbf{S}} + \underbrace {\left( {{\mathbf{HX}} - {\mathbf{S}}} \right)}_{{\text{MUI}}} + {\mathbf{W}}_C,
\end{equation}
where each entry of $\mathbf{S}$ is randomly drawn from a given constellation. The second term at the right-hand side of (\ref{eq2}) can be interpreted as the MUI that interfere the symbol detection in an AWGN channel, with its total energy being defined as
\begin{equation}\label{eq3}
  {P_{{\text{MUI}}}} = \left\| {{\mathbf{HX}} - {\mathbf{S}}} \right\|_F^2.
\end{equation}
It has been shown in \cite{MUI_sum_rate} that by minimizing the MUI, the lower-bound of the achievable sum-rate can be maximized. In the remainder of the paper, we will employ (\ref{eq3}) as a performance metric for downlink communications.
\subsection{Radar Model}
Let us consider a single target of interest located in range bin $p = 0$ and angle $\theta_0$, surrounded by $M$ unwanted scatterers located at angles $\theta_m$ within a collection of range bins $\left\{ { - P,..., - 1,1,...,P} \right\}$, where $P$ is the largest range bin of interest. The target echo can be therefore given as \cite{6384816}
\begin{equation}\label{eq4}
\begin{gathered}
  {{\mathbf{Y}}_R} = {\alpha _0}{\mathbf{a}}\left( {{\theta _0}} \right){{\mathbf{a}}^H}\left( {{\theta _0}} \right){\mathbf{X}} \hfill \\
   + \sum\limits_{p =  - P,p \ne 0}^P {\sum\limits_{m = 1}^M {{\alpha _{p,m}}{\mathbf{a}}\left( {{\theta _m}} \right){{\mathbf{a}}^H}\left( {{\theta _m}} \right){\mathbf{X}}} }{{{\mathbf{J}}_p}}  + {{\mathbf{W}}_R}, \hfill \\
\end{gathered}
\end{equation}
where ${\mathbf{a}}\left( {{\theta}} \right)$ represents the ULA steering vector, $\alpha_{0}$ and $\alpha_{p,m}$  denote the complex amplitudes proportional to the radar cross-section (RCS) of the targets and the scatterers, respectively, ${\mathbf{W}}_R$ is the noise matrix, and $\mathbf{J}_p $ is defined as the following temporal shifting matrix
\begin{equation}\label{eq5}
  {{\mathbf{J}}_p} = \left[ {\begin{array}{*{20}{c}}
  {\underbrace {0 \ldots 0}_{\text{p zeros}}}&1& \ldots &0 \\
  {0 \ldots 0}&0& \ddots & \vdots  \\
   \vdots & \vdots & \vdots &\begin{gathered}
  1 \hfill \\
   \vdots  \hfill \\
\end{gathered}  \\
  {0 \ldots 0}&0& \ldots &0
\end{array}} \right] \in {\mathbb{R}^{L \times L}} .
\end{equation}
It follows that $\mathbf{J}_p = \mathbf{J}_{-p}^T$ and $\mathbf{J}_0 = \mathbf{I}_L$, where $\mathbf{I}_L$ is the size-\emph{L} identity matrix. After range compression, the output signal of the matched filter can be obtained by
\begin{equation}\label{eq6}
\begin{gathered}
  {\mathbf{D}} = \frac{1}{L}{{\mathbf{Y}}_R}{{\mathbf{X}}^H} = \frac{1}{L}{\alpha _0}{\mathbf{a}}\left( {{\theta _0}} \right){{\mathbf{a}}^H}\left( {{\theta _0}} \right){\mathbf{X}}{{\mathbf{X}}^H} \hfill \\
   + \frac{1}{L}\sum\limits_{p =  - P,p \ne 0}^P {\sum\limits_{m = 1}^M {{\alpha _{p,m}}{\mathbf{a}}\left( {{\theta _m}} \right){{\mathbf{a}}^H}\left( {{\theta _m}} \right){\mathbf{X}}{{\mathbf{J}}_p}} } {{\mathbf{X}}^H} + {{\mathbf{W}}_R}{{\mathbf{X}}^H}, \hfill \\
\end{gathered}
\end{equation}
where the second term in (\ref{eq6}) denotes the clutter interference which needs to be reduced. Let us define the integrated range sidelobe power as
\begin{equation}\label{eq7}
{P_\text{ISL}} = \sum\limits_{p = -P, p \ne 0}^P {\left\| {{\mathbf{X}}{{\mathbf{J}}_p}{{\mathbf{X}}^H}} \right\|_F^2}.
\end{equation}
Given the DFRC waveform matrix $\mathbf{X}$, the MIMO radar transmit beampattern is defined as \cite{li2008mimo}
\begin{equation}\label{eq8}
  G\left( \theta  \right) = {{\mathbf{a}}^H}\left( \theta  \right){{\mathbf{R}}_X}{\mathbf{a}}\left( \theta  \right),
\end{equation}
where ${{\mathbf{R}}_X} = \frac{1}{L}{\mathbf{X}}{{\mathbf{X}}^H}$ is the waveform covariance matrix.

\section{Problem Formulation}
In this section, we firstly recall the closed-form design in \cite{8386661}, and then formulate the optimization problems for DFRC waveform design based on the radar and the communication models above.
\subsection{Closed-form Design for Given Radar Beampatterns}
The closed-form design in \cite{8386661} without considering the sidelobe minimization is formulated as
\begin{equation}\label{eq9}
  \mathop {\min }\limits_{\mathbf{X}} \;P_{\text{MUI}}\;s.t.\;{{\mathbf{R}}_X} = {{\mathbf{R}}_d},
\end{equation}
where ${{\mathbf{R}}_d}$ is a given covariance matrix corresponds to a well-designed beampattern. In (\ref{eq9}), the communication MUI is minimized under an equality constraint that guarantees that the desired beampattern is achievable. While problem (\ref{eq9}) is non-convex, a globally optimal solution can be readily obtained in closed-form, which is
\begin{equation}\label{eq10}
  {\mathbf{X}} = \sqrt L {\mathbf{FU}}{{\mathbf{I}}_{N \times L}}{{\mathbf{V}}^H},
\end{equation}
where ${\mathbf{F}}{{\mathbf{F}}^H} = {{\mathbf{R}}_d}$ is the Cholesky decomposition of ${{\mathbf{R}}_d}$, and ${\mathbf{U\Sigma }}{{\mathbf{V}}^H} = {{\mathbf{F}}^H}{{\mathbf{H}}^H}{\mathbf{S}}$ denotes the SVD of ${{\mathbf{F}}^H}{{\mathbf{H}}^H}{\mathbf{S}}$. We refer readers to \cite{8386661} for a detailed derivation of the solution (\ref{eq10}).

\subsection{Radar-Communication Trade-off Design}
While the solution of (\ref{eq9}) can guarantee a desired beampattern, it cannot suppress the range sidelobe level. More importantly, the strong constraint in (\ref{eq9}) may lead to serious performance-loss in downlink communication. We therefore define a relaxed waveform similarity metric by relying on the following squared Euclidean distance
\begin{equation}\label{eq11}
{P_{{\text{SIM}}}} = \left\| {{\mathbf{X}} - {{\mathbf{X}}_0}} \right\|_F^2,
\end{equation}
where ${\mathbf{X}}_0$ is a reference waveform matrix obtained from solving (\ref{eq9}). By minimizing (\ref{eq11}), one can approach the desired radar beampattern without imposing an equality constraint on the waveform covariance matrix.  Accordingly, the trade-off optimization problem for DFRC waveform design can be formulated as
\begin{equation}\label{eq12}
\begin{gathered}
  \mathop {\min }\limits_{\mathbf{X}} \;F\left( {\mathbf{X}} \right) = {\rho _1}{P_{{\text{MUI}}}} + {\rho _2}{P_{{\text{ISL}}}} + {\rho _3}{P_{{\text{SIM}}}} \hfill \\
  s.t.\;\;\operatorname{diag} \left( {{{\mathbf{R}}_X}} \right) = \frac{{{P_T}}}{N}{{\mathbf{1}}_N}, \hfill \\
\end{gathered}
\end{equation}
where ${P_T}$ is the total transmit power budget, ${\mathbf{1}}_N$ denotes a size-\emph{N} all-one vector, and $\operatorname{diag} \left( {\mathbf{R}}_X\right)$ is a vector composed by the diagonal entries of $\mathbf{R}_X$, which represents the transmit power at each antenna. We impose an equality diagonal constraint here due to the facts that the radar transmits at its maximum available power budget, and that it typically requires a per-antenna power control. By formulating (\ref{eq12}), we aim at minimizing the weighted summation of $P_{\text{MUI}}$, $P_{\text{ISL}}$ and ${P_{{\text{SIM}}}}$, with $\rho_i \ge 0, \forall i$ being the weighting factors that indicate the priority of the three performance metrics.
\\\indent Given the non-convexity in $P_{\text{SIM}}$ as well as in the equality power constraint, problem (\ref{eq12}) can not be easily solved via convex optimization algorithms. In what follows, we will present a first-order Riemannian Gradient Conjugate (RCG) algorithm \cite{mani_optimization} for solving the problem based on the complex oblique manifold.
\section{Proposed Algorithm based on Oblique Manifold}
It is straightforwardly to see that
\begin{equation}\label{eq13}
  {\rho _1}{P_{{\text{MUI}}}} + {\rho _2}{P_{{\text{ISL}}}} = \left\| {{\mathbf{AX}} - {\mathbf{B}}} \right\|_F^2,
\end{equation}
where
\begin{equation}\label{eq14}
  {\mathbf{A}} = \left[ \begin{gathered}
  \sqrt {{\rho _1}} {\mathbf{H}} \hfill \\
  \sqrt {{\rho _2}} {{\mathbf{I}}_N} \hfill \\
\end{gathered}  \right],{\mathbf{B}} = \left[ \begin{gathered}
  \sqrt {{\rho _1}} {\mathbf{S}} \hfill \\
  \sqrt {{\rho _2}} {{\mathbf{X}}_0} \hfill \\
\end{gathered}  \right].
\end{equation}
By the above notations, problem (\ref{eq12}) can be recast as
\begin{equation}\label{eq15}
\begin{gathered}
  \mathop {\min }\limits_{\mathbf{X}} \;F\left( {\mathbf{X}} \right) = \left\| {{\mathbf{AX}} - {\mathbf{B}}} \right\|_F^2 + {\rho _3}\sum\limits_{p =  - P,p \ne 0}^P {\left\| {{\mathbf{X}}{{\mathbf{J}}_p}{{\mathbf{X}}^H}} \right\|_F^2}  \hfill \\
  s.t.\;\;\operatorname{diag} \left( {{\mathbf{X}}{{\mathbf{X}}^H}} \right) = \frac{{L{P_T}}}{N}{{\mathbf{1}}_N}. \hfill \\
\end{gathered}
\end{equation}
By taking a closer look at the power constraint, we note that the feasible region of problem (\ref{eq15}) forms an \emph{NL}-dimensional complex \emph{oblique manifold} \cite{mani_optimization}, which is a Riemannian manifold. To accelerate the process of solving the problem, the iterative algorithm should operate along the descent direction on the manifold rather than on the ambient Euclidean space. Typically, such directions can be found on the so-called \emph{tangent space}. Let $\mathcal{M}$ be the feasible region of (\ref{eq15}). Given a point $\mathbf{X} \in \mathcal{M}$, i.e., a feasible solution to problem (\ref{eq15}), the tangent space is defined as the set of all the tangent vectors that are tangential to any smooth curves on $\mathcal{M}$ through $\mathbf{X}$. This can be mathematically expressed as
\begin{equation}\label{eq16}
{T_{\mathbf{X}}}\mathcal{M} = \left\{ {{\mathbf{Z}} \in {\mathbb{C}^{N \times L}}\left| {\operatorname{Re} \left( {{{\left( {{{\mathbf{X}}^H}{\mathbf{Z}}} \right)}_{ii}}} \right) = 0,\forall i} \right.} \right\}.
\end{equation}
To proceed the RCG algorithm, we first calculate the gradient of the objective function $F\left(\mathbf{X}\right)$ as follows
\begin{equation}\label{eq17}
\begin{gathered}
  \nabla F\left( {\mathbf{X}} \right) = 2{{\mathbf{A}}^H}\left( {{\mathbf{AX}} - {\mathbf{B}}} \right) \hfill \\
  {\text{ + }}\sum\limits_{p =  - P,p \ne 0}^P {2\left( {{\mathbf{X}}{{\mathbf{J}}_p}{{\mathbf{X}}^H}{\mathbf{XJ}}_p^H + {\mathbf{XJ}}_p^H{{\mathbf{X}}^H}{\mathbf{X}}{{\mathbf{J}}_p}} \right)}.  \hfill \\
\end{gathered}
\end{equation}
One of the key step in RCG is to project the above \emph{Euclidean gradient} (\ref{eq17}) onto the tangent space (\ref{eq16}), which yields an ascent direction on the manifold as
\begin{equation}\label{eq18}
\begin{gathered}
  \operatorname{grad} F\left( {\mathbf{X}} \right) = {\mathcal{P}_{\mathbf{X}}}\left(\nabla F\left( {\mathbf{X}} \right)\right) \hfill \\
   = \nabla F\left( {\mathbf{X}} \right) - {\mathbf{X}}\operatorname{ddiag} \left( {{{\mathbf{X}}^H}\nabla F\left( {\mathbf{X}} \right)} \right), \hfill \\
\end{gathered}
\end{equation}
where the operator $\operatorname{ddiag}\left(\cdot\right)$ sets all the off-diagonal entries of the input matrix as zero. Eq. (\ref{eq18}) is named as the \emph{Riemannian gradient} in contrast to its Euclidean counterpart (\ref{eq17}). We next compute the conjugate descent direction ${{\mathbf{\Pi }}_k}$ at the \emph{k}-th iteration. Recall that in the classic conjugate gradient method, the descent direction at the \emph{k}-th iteration is a linear combination of the \emph{k}-th gradient and the $\left(k-1\right)$-th descent direction \cite{nocedal2006numerical}. In the RCG framework, however, $\operatorname{grad} F\left( {{{\mathbf{X}}_k}} \right)$ and ${{{\mathbf{\Pi }}_{k - 1}}}$ belong to different tangent spaces, which can not be linearly combined. As such, we firstly project ${{{\mathbf{\Pi }}_{k - 1}}}$ onto ${T_{{{\mathbf{X}}_k}}}\mathcal{M}$, and then combine it with the associated negative Riemannian gradient as \cite{mani_optimization}
\begin{equation}\label{eq19}
{{\mathbf{\Pi }}_k} =  - \operatorname{grad} F\left( {{{\mathbf{X}}_k}} \right) + {\lambda _k}{\mathcal{P}_{{{\mathbf{X}}_k}}}\left( {{{\mathbf{\Pi }}_{k - 1}}} \right),
\end{equation}
where $\lambda_k$ is the Polak-Ribi\'ere combination coefficient, which can be obtained as
\begin{equation}\label{eq20}
{\lambda _k} = \frac{{\left\langle {\operatorname{grad} F\left( {{{\mathbf{X}}_k}} \right),\operatorname{grad} F\left( {{{\mathbf{X}}_k}} \right) - {\mathcal{P}_{{{\mathbf{X}}_k}}}\left( {F\left( {{{\mathbf{X}}_{k - 1}}} \right)} \right)} \right\rangle }}{{\left\langle {\operatorname{grad} F\left( {{{\mathbf{X}}_{k - 1}}} \right),\operatorname{grad} F\left( {{{\mathbf{X}}_{k - 1}}} \right)} \right\rangle }},
\end{equation}
where $\left\langle \cdot,\cdot  \right\rangle$ represents the matrix inner product. However, by moving towards the direction of ${{\mathbf{\Pi }}_k}$, the resultant point is still on the tangent space ${T_{{{\mathbf{X}}_k}}}\mathcal{M}$ rather than on the manifold $\mathcal{M}$. Therefore, the following \emph{Retraction mapping} is defined to retract a point on ${T_{{{\mathbf{X}}_k}}}\mathcal{M}$ to its nearest neighbor on $\mathcal{M}$  \cite{mani_optimization}
\begin{equation}\label{eq21}
  {\mathcal{R}_{\mathbf{X}}}\left( {\mathbf{Z}} \right) = \beta \operatorname{ddiag} {\left( {\left( {{\mathbf{X}} + {\mathbf{Z}}} \right){{\left( {{\mathbf{X}} + {\mathbf{Z}}} \right)}^H}} \right)^{ - \frac{1}{2}}}\left( {{\mathbf{X}} + {\mathbf{Z}}} \right),
\end{equation}
where $\beta  = \sqrt {\frac{{L{P_T}}}{N}}$ is a scaling factor.
\\\indent Based on the above principle, we are now ready to present the RCG method, which is summarized in Algorithm 1.
 \begin{figure*}[!t]
\centering
\subfloat[]{\includegraphics[width=1.85in]{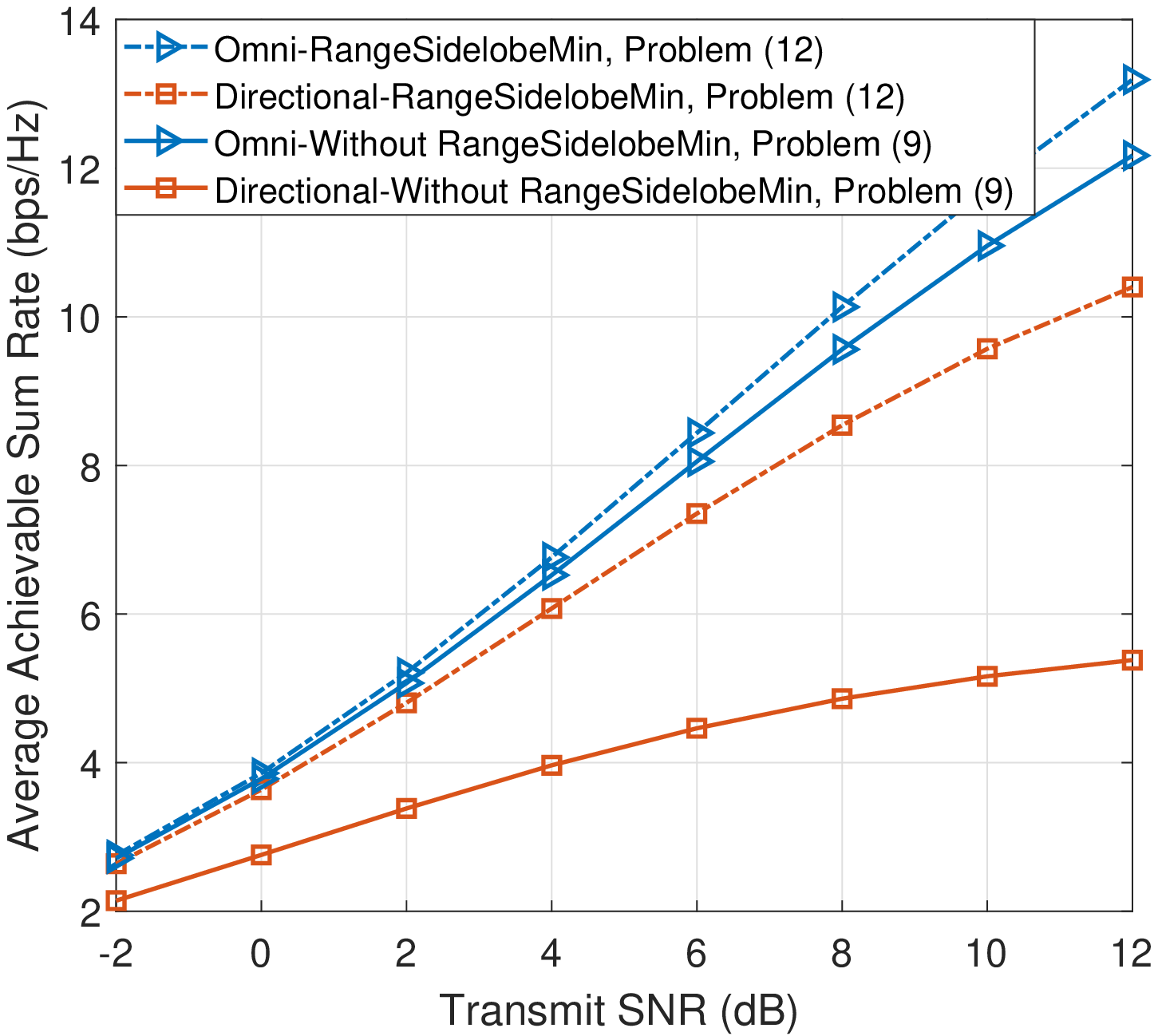}
\label{fig2}}
\hspace{.01in}
\subfloat[]{\includegraphics[width=1.85in]{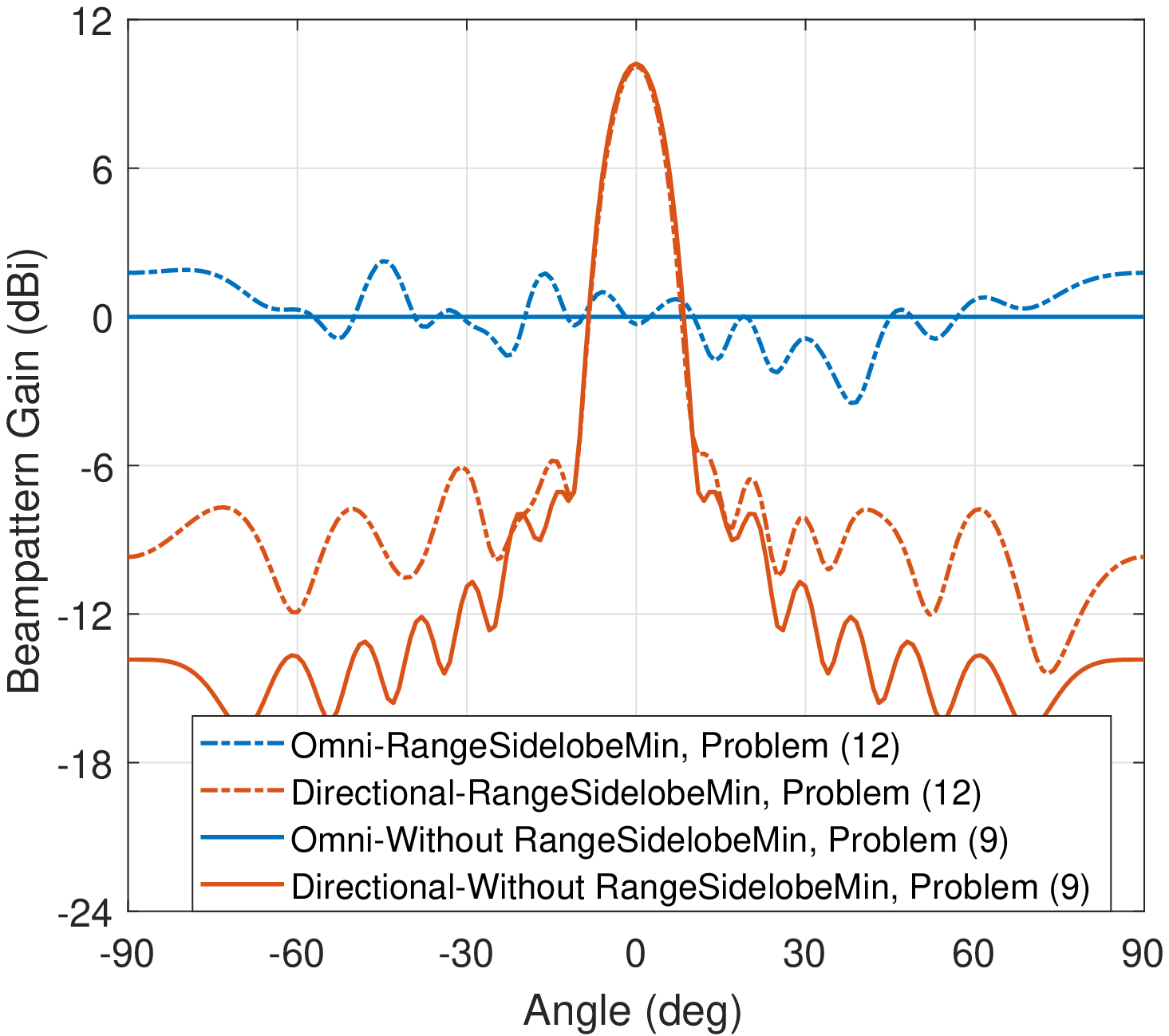}
\label{fig3}}
\hspace{.01in}
\subfloat[]{\includegraphics[width=1.85in]{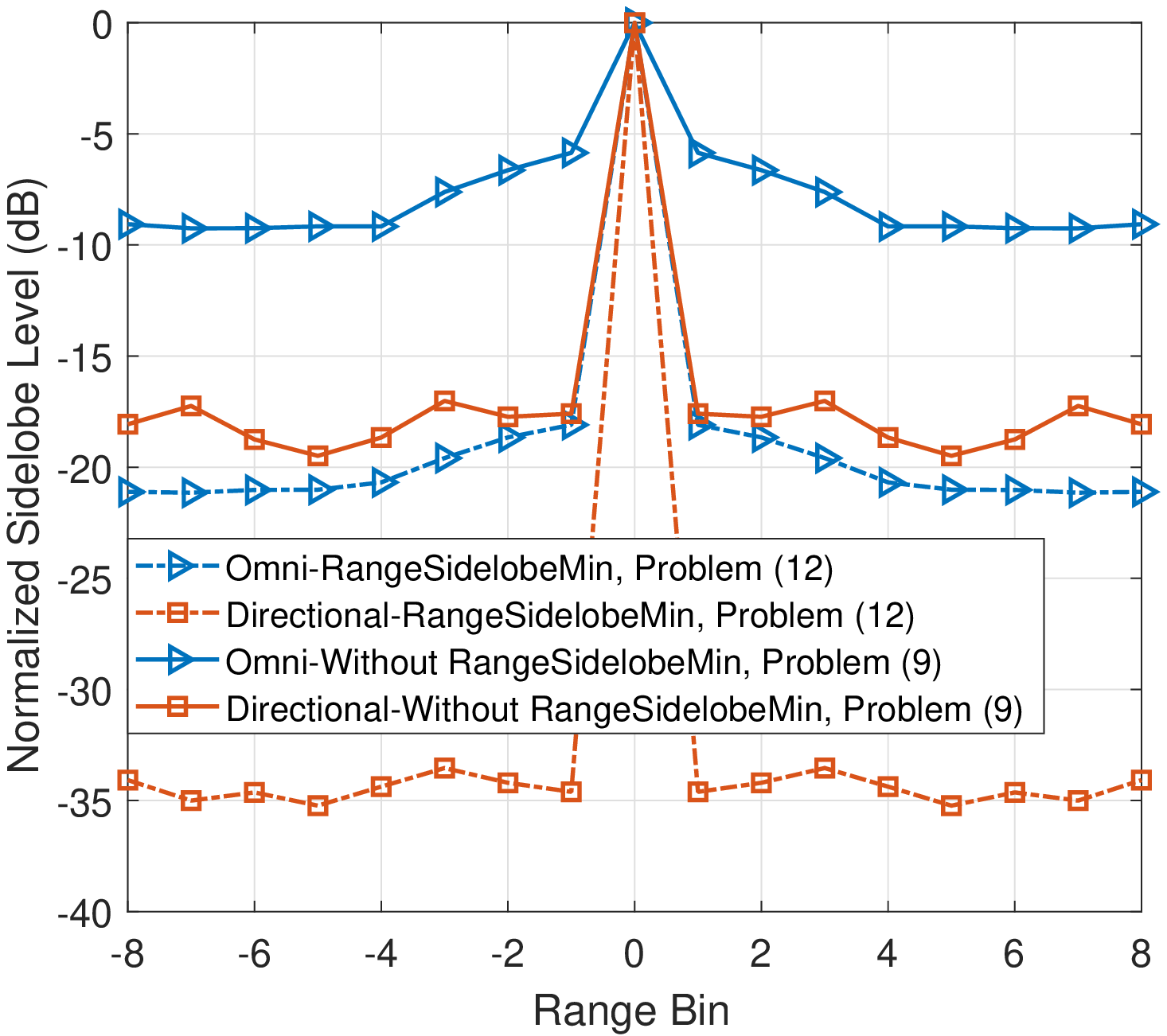}
\label{fig4}}
\caption{Numerical results. (a) Average achievable sum-rate for different methods; (b) Radar spatial beampattern for different methods; (c) Radar range sidelobe level for different methods.}
\label{fig_sim}
\end{figure*}
\\\indent \emph{Remark:} The computational complexity of Algorithm 1 is dominated by the calculation of the Euclidean gradient (\ref{eq17}), which requires $\mathcal{O}\left( {{N^2}PL - {{{N^2}{P^2}} \mathord{\left/
 {\vphantom {{{N^2}{P^2}} 2}} \right.
 \kern-\nulldelimiterspace} 2}} \right)$ complex multiplications per iteration. As the strict convergence analysis of the RCG approach still remains open problem \cite{mani_optimization}, it is difficult to specify the iteration number needed for Algorithm 1. Nevertheless, we observe in our simulations that the algorithm converges within tens of iterations for a tolerable gradient norm of $\varepsilon = 10^{-6}$.
\renewcommand{\algorithmicrequire}{\textbf{Input:}}
\renewcommand{\algorithmicensure}{\textbf{Output:}}
\begin{algorithm}
\caption{RCG Algorithm for Solving (\ref{eq15})}
\label{alg:C}
\begin{algorithmic}
    \REQUIRE $\mathbf{H},\mathbf{S},{\mathbf{X}}_0$, weighting factor $\rho_i, \forall i$, $P_T$, the largest range bin of interest $P$, tolerable error $\varepsilon > 0$, maximum iteration number $k_{max} > 2$
    \ENSURE $\mathbf{X}_{k+1}$
    \STATE 1. Compute $\mathbf A$, $\mathbf B$ via (\ref{eq14}). Initialize randomly $\mathbf{X}_{0} = \mathbf{X}_{1} \in \mathcal{M}$, set $\mathbf{\Pi}_0=-\operatorname{grad}F\left(\mathbf{X}_{0}\right)$, $k=1$.
    \WHILE{$k\le k_{max}$ and ${\left\| {\operatorname{grad} F\left( {{{{\mathbf{X}}}_{k}}} \right)} \right\|_F} \ge {\varepsilon}$}
    \STATE 2. Compute the combination coefficient $\lambda_k$ by (\ref{eq20}).
    \STATE 3. Compute the descent direction ${\mathbf{\Pi} _k}$ by (\ref{eq19}).
    \STATE 4. Compute stepsize $\mu_{k}$ by the Armijo line search method, and set ${{{\mathbf{X}}}_{k+1}}$ by
    \begin{equation*}
    {{\mathbf{X}}_{k+1}} = {\mathcal{R}_{{{\mathbf{X}}_{k}}}}\left( {{\mu_k}{{\mathbf{\Pi }}_k}} \right).
    \end{equation*}
    \STATE 5. $k = k + 1.$
    \ENDWHILE
\end{algorithmic}
\end{algorithm}

\section{Numerical Results}
In this section, we assess the performance of the proposed waveform design (\ref{eq12}) by Monte-Carlo simulations. Without loss of generality, we set $N = 16$, $K = 4$, $P_T = 1$, $L = 100$ and $P = 8$. The communication channel is assumed to be Rayleigh fading, where each entry of $\mathbf{H}$ subjects to standard complex Gaussian distribution. The communication data matrix $\mathbf{S}$ is comprised by unit-power QPSK symbols. For completeness, we consider both omni-directional and directional waveform designs for the radar functionality. In the first design, the desired covariance matrix is given as ${{\mathbf{R}}_d} = \frac{{{P_T}}}{N}{{\mathbf{{\rm I}}}_N}$, which results in an omni-directional beampattern. In the second design, on the contrary, $\mathbf{R}_d$ is generated following the method of [Eq. (1.93), 5], which leads to a directional beampattern focusing on $0^\circ$ with a 3dB beamwidth of $10^\circ$. For comparison, the closed-form design (\ref{eq9}) is employed as the benchmark, where the range sidelobe reduction is not addressed. In all the results, the weighting factors for (\ref{eq12}) are set as $\rho_1 = \rho_3 = 0.15$ and $\rho_2 = 0.7$.
\\\indent We first look at the achievable sum-rate of the downlink users and the corresponding radar beampatterns as shown in Fig. 2(a)-(b). The sum-rates are computed based on [Eq. (5), 4]. The solid and the dashed lines represent the performance of the closed-form and the proposed waveform designs (\ref{eq9}) and (\ref{eq12}), respectively, where we see that by introducing only a small weighting factor $\rho_1 = 0.15$ to the communication functionality, the sum-rate performance increases significantly. In the meantime, we observe small mismatches in Fig. 2(b) between the resultant spatial beampatterns from solving (\ref{eq12}) and their reference counterparts of solving (\ref{eq9}). However, such performance-loss is acceptable given the considerable gain obtained in the communication rate. In Fig. 2(c), we further demonstrate the performances of both the benchmark and the proposed designs in terms of the normalized range sidelobe level. It is noteworthy that by solving (\ref{eq12}) under only a small weighting factor $\rho_3 = 0.15$, we obtain a 12dB sidelobe reduction in the omni-directional waveform design, and a 17dB reduction in its directional counterpart, which again proves the superiority of the proposed method.
\section{Conclusion}
In this paper, we have proposed a novel waveform design for DFRC systems by minimizing the weighted summation of the multi-user communication interference, the Euclidean distance between the designed and the reference waveforms, and the integrated range sidelobe level. To solve the non-convex optimization problem formulated, we have proposed an efficient algorithm based on the oblique manifold. Finally, we have demonstrated by numerical simulations that the proposed method significantly outperforms the benchmark closed-form design \cite{8386661} in both the communication sum-rate and the range sidelobe reduction, with an acceptable mismatch in the formulated spatial beampatterns for the radar functionality.


%




\ifCLASSOPTIONcaptionsoff
  \newpage
\fi


\bibliographystyle{IEEEtran}
\bibliography{IEEEabrv,RadCom}
\end{document}